\documentclass[%
 reprint,
 amsmath,amssymb,
 aps,
]{revtex4-1}

\usepackage{graphicx}
\usepackage{dcolumn}
\usepackage{bm}
\usepackage[mathlines]{lineno}
\usepackage{braket}
\usepackage{hyperref}

\begin{document}

\preprint{}

\title{Azimuthal angular entanglement between decaying particles in ultra-peripheral ion collisions}

\author{Spencer R. Klein}
 \email{srklein@lbl.gov}
\affiliation{
 Lawrence Berkeley National Laboratory, Berkeley, CA, USA}

\date{\today}
\begin{abstract}

Ultra-peripheral collisions (UPCs) involving relativistic heavy ions are a unique laboratory to study quantum correlations. The intense electromagnetic fields generate high rates of photonuclear interactions, including events involving multiple photon exchange.   Multiple photon exchange can result in the production of multiple vector mesons and/or nuclear excitations.  These interactions share a common impact parameter, so the photons have the same linear polarization.  The shared polarization entangles the particles, leading to unique quantum correlations.  The decays of these vector excitations are sensitive to this polarization, allowing for the study of these correlations.

This letter will compare classical and quantum calculations of the correlations between these azimuthal directions.  The two approaches predict vert different angular correlations.  The differences are akin to those seen with polarized photons in tests of Bells inequality.  Uniquely, UPC photoproduction can produce final states containing three or more particles, all entangled with the same polarization.  These more complex states exhibit additional unique phenomenology, allowing new tests of multi-particle entanglement. 

\end{abstract}

\maketitle

Ultra-peripheral collisions (UPCs) of relativistic heavy ions are photon-mediated reactions that occur when two ions interact at a minimum separation (impact parameter, $|\vec{b}|$) large enough so that no hadronic interactions occur- typically $|\vec{b}|\gtrapprox 15$\ fm for lead or gold \cite{Bertulani:2005ru,Baltz:2007kq,Klein:2020fmr}. Either ion can be a photon source or target.  In exclusive photoproduction of vector mesons, the two sources act like a two-source interferometer. The amplitudes for production on each target add.  Because of the opposite $\vec{E}$ directions, the two amplitudes have opposite signs, so there is destructive interference when $p_T < \hbar/\langle b\rangle$ \cite{Klein:1999gv,STAR:2008llz,Brandenburg:2025one}. 
The phase-coherence of this interferometer comes from the initial-state symmetry, since the two amplitudes have no causal connection, so are an example of a non-local system \cite{Klein:2002gc,Brandenburg:2025one}. This interference also exhibits an angular modulation, depending on the azimuthal angle between $\vec{b}$ and the direction of the decay pions \cite{Xing:2020hwh,STAR:2022wfe,ALICE:2024ife}.

The electromagnetic fields from heavy ion are very intense, and a single ion-ion encounter may involve the exchange of multiple photons, producing multiple final state particles.  Not surprisingly, multiple particles exhibit more complex interference phenomena \cite{Klein:2024wos}.  This letter will discuss the polarization entanglement since all of produced photons share a common polarization vector.  These systems have similarities to those used to study Bell's inequality, albeit with self-analyzing, decaying particles.   

The nuclear electric fields extend radially outward, so at the position of the other ion, $\vec{b}\parallel \vec{E}$, and all of the emitted photons are linearly polarized in the same direction (or 180$^{\circ}$ out of phase).  Each photon is emitted independently, save for the common $\vec{b}$ \cite{Gupta:1955zza}.

This situation has parallels with experiments using pairs of entangled photons to test Bell's inequality.  Those studies measured the joint probabilities for photons to pass through filters with adjustable relative orientations \cite{Aspect:1982fx}. A photon passing through a filter acquires the orientation of the filter, and any entangled photons will follow suit. Bell showed that quantum mechanical expectations are very different from any classical model \cite{Bell:1964kc}.  With UPCs, instead of using filters, the particle decays are self-analyzing.  The polarization is determined by measuring the azimuthal directions of the daughter particles.
 
Polarization  correlations using self-analysis have been previously studied in reactions such as $e^+e^-\rightarrow\tau^+\tau^-$ \cite{Privitera:1991nz}, $\phi\rightarrow K^0\overline K^0$ \cite{Bertlmann:2001ea}, $e^+e^-\rightarrow\Lambda\overline\Lambda$  \cite{Tornqvist:1980af} and $\eta_c\rightarrow VV$ \cite{Li:2009rta,Chen:2012xa}, where $V$ is a VM.  Unlike UPC interferometers, all of these pairs have a common point of origin.  

\begin{figure}
\center{\includegraphics[width=0.9\linewidth]{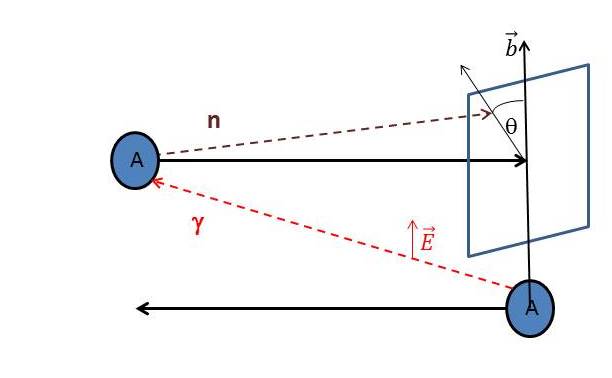}}
\center{\includegraphics[width=0.9\linewidth]{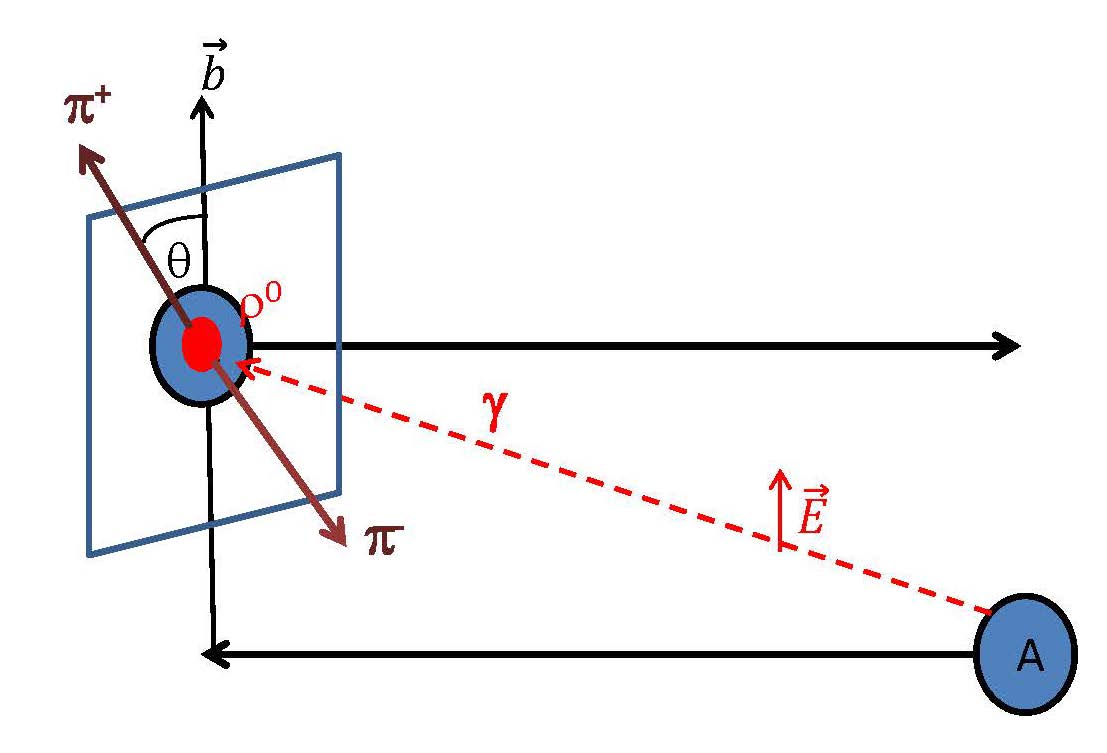}}
\caption{(top) The geometry for UPC photoexcitation to a giant dipole resonance (GDR) state.  At closest approach, the two nuclei (``A") are separated by $\vec{b}$.  One of them emits a photon $\gamma$ with polarization $\vec{E}$ parallel to $\vec{b}$ which strikes the other nucleus, exciting it to a GDR, which then decays, emitting a single neutron (``n").  Perpendicular to the beam direction, the neutron makes an angle $\theta$ with respect to $\vec{b}$. (bottom) The geometry for UPC photoproduction of a $\rho^0$ (in red), almost immediately followed by its decay to $\pi^+\pi^-$.
\label{fig:geometry}}
\end{figure}

In UPCs two types of interactions are possible.  Although the types are very different, the polarization correlations should be the same. In the first class, a photon from one nucleus can excite the other nucleus.  Often, this is a short-lived excitation known as a Giant Dipole Resonance (GDR), which usually decays by emitting a single neutron \cite{Berman:1975tt}.  Figure \ref{fig:geometry} shows the geometry for this decay.  The excited nucleus retains the photon polarization; the neutron is emitted at an azimuthal angle $\theta$ with respect to the photon polarization.   At heavy ion colliders like RHIC and the LHC, these neutrons may be detected in position-sensitive zero degree calorimeters \cite{Snote}.  

Alternately, in coherent photoproduction, photons  fluctuate to $q\overline q$ dipoles which then scatter elastically from the target, emerging as vector mesons (VMs).  Normally, `coherent' means that the target nucleus remains intact, although coherent addition of amplitudes has been observed even when the target nuclei dissociate \cite{Klein:2023zlf}.  VMs have the same quantum numbers as photons, $J^{PC} = 1 ^{--}$. Per s-channel helicity conservation (SCHC) the vector mesons have the same polarization as the incident photon \cite{Bauer:1977iq,HERAdata}.  SCHC has been experimentally verified for both GDR excitation \cite{Berman:1975tt} and VM photoproduction \cite{Ballam:1970ex,STAR:2007elq,ALICE:2023svb}.   The vector state retains the linear polarization of the incident photon to within a few percent.

VMs decay quickly, with lifetimes less than $10^{-19}$ s.  The most commonly studied final states involve two daughters, such as in $\rho^0\rightarrow\pi^+\pi^-$, $\phi\rightarrow K^+K^-$ or $J/\psi\rightarrow e^+e^-$.  The daughter particles can be detected in tracking chambers.  In these decays,  the azimuthal angle between $\vec{b}$ and the 
daughters ({\it i.e.} of the $\pi^+$) is distributed following \cite{Baur:2003ar}
\begin{equation}
P(\theta) \approx \cos^2(\theta)
\label{eq:DTone}
\end{equation}
(cf. Fig. 1).  In GDR decays, the neutron transverse direction of the neutron follows the same distribution
\cite{Bertulani:1987tz}.

If the photons or vector mesons had a significant transverse (to the beam direction) momentum then that might affect the observed correlations.  The typical photon $p_T$, is small, typically of order $\hbar/\langle |\vec{b}|\rangle \lesssim 10$ MeV/c.  For coherent vector meson (VM) production, one must also add (in quadrature) the momentum transfer from the scattering. This is roughly a few $\hbar/R_A\approx 30$ MeV/c, where $R_A$ is the nuclear radius.  These $p_T$ are small enough so that the transverse boosts do not significantly affect the angular correlations discussed here.

There are two ways to produce two particles. Either one nucleus can emit both photons, or each can emit one photon.   In the latter case, both the emission and the photon-target interactions are manifestly independent, except for the common $\vec{b}$.   

Even when both photons come from a single nucleus, photon emission is independent, as long as the photon energy is small compared to the ion energy \cite{Gupta:1955zza}.  When multiple photons excite a single target nucleus, this can lead to a more energetic excitation, usually involving multiple neutrons.  We will not consider this case.  Coherent VM photoproduction involves elastic scattering, which generally leaves the target unchanged, in the ground state, so, in multiple VM production, production of multiple mesons on the same target should be independent \cite{Baur:2003ar} - except for the identical polarization. 

The cross-section for single or multiple UPC photoproduction may be calculated using the Weizs\"acker-Williams approach, whereby the electromagnetic field of one nucleus is represented as a field of virtual photons \cite{Bertulani:2005ru,Baltz:2007kq,Klein:2020fmr}. 

Because the interactions are independent, the probability to produce a complex final state $x_1$, $x_2$,... in an encounter at impact parameter $b$ is
\begin{equation}
P(x_1,x_2...) \int d^2b P_{x_1}(b) P_{x_2}(b)....(1-P_{\rm had}(b))
\end{equation}
where $P_{\rm had}$ is the probability for a hadronic interaction which will obscure the photon-induced interactions.  For coherent production from heavy ions, the probability of a single photon doing two things (such as producing a VM and exciting the emitting nucleus) is small \cite{Baur:2003ar}.  
This cross section is the probability, integrated over all possible $\vec{b}$. 

This independent-interaction concept is sometimes known as factorization.   The measured cross sections for VM production accompanied by single or multiple Coulomb excitation \cite{STAR:2002caw,STAR:2007elq,ALICE:2020ugp} agree with calculations which assume factorization \cite{Baltz:2002pp}.  Despite the multi-photon exchange, these  cross sections are large. For example, the cross-sections for mutual GDR excitation at the LHC is about 550 mb \cite{Baltz:1998ex}.  

We  now consider classical and quantum mechanical calculations of the angular correlations between the final states.   We consider only the entangled polarizations, and neglect correlations that occur for identical particle types.   We also assume that all of the vector particles are distinguishable, ({\it e. g.} final states like $\rho\phi J\psi$ or mutual Coulomb excitation accompanied by a VM). 

{\it Classical calculation.} Classically, each produced vector particle follows Eq. \ref{eq:DTone} with respect to $\vec{b}$.   For two vector excitations, the azimuthal angle between two of the decay products is given by the convolution of the two  $\cos^2(\theta)$ distributions, so the azimuthal angular separation between the two neutrons, $\theta_{21}$ follows \cite{Baur:2003ar} 
\begin{equation}
P(\theta_{21}) = \int d\theta P(\theta_1-\phi) P(\theta_2-\theta)\delta(\Delta\theta-(\theta_1-\theta_2))
\end{equation}
where $\delta$ is the Kronecker delta function.  Either decay product can be used, since the angular distribution is symmetric under 180$^\circ$ rotation.

Simplifying leads to the angular correlation
\begin{equation}
P(\theta_{21}) =  1 + \frac{\cos(2\theta_{21})}{2}
\label{eq:DTclass}
\end{equation}
If additional vector particles are emitted, they will also follow the $\cos^2(\theta)$ distribution with respect to $\vec{b}$ and so Eq. \ref{eq:DTclass} applies for every pair of vector particles emitted in the collision.

{\it Quantum correlations.} Quantum mechanically, $\vec{b}$ is not directly observable, so the initial amplitude is the same for all possible azimuthal directions. The photon polarizations are likewise undetermined. All of the exchanged photons share the same polarization, so the vector particles must have parallel (for photons from the same nucleus) or anti-parallel (for photons emitted by opposing nuclei) polarization.  It is possible to select kinematic regions to emphasize aligned or anti-aligned polarization, but that distinction is unimportant here.  Normally,  the two possible directions for the spin-1 particle are labeled up/down, or sideways, for polarization in the vertical plane, or polarization in the horizontal plane.  Then, the wave function for a two vector particle state is
$
\ket{\psi_2} = \ket{V_1V_2} = \frac{1}{\sqrt 2} \ket{\uparrow_1 \uparrow_2} + \frac{1}{\sqrt 2} \ket{\leftarrow_1 \leftarrow_2} 
$.
This formulation, however, neglects the presence of the two ions and the azimuthal angle $\theta$ for their impact parameter vector.  The ions have high momentum, so $\theta$ is not quantized. Instead, we can write
\begin{equation}
 \ket{\psi_2} = \ket{V_1(\theta) V_2 (\theta)b(\theta)} 
\end{equation}
where $\theta$ is the azimuthal angle.  $\theta$ is randomly distributed, so the wave function is azimuthally symmetric, with no preferred azimuthal direction.   The photons are polarized along angle $\theta$. This wave function cannot be factorized into independent single-particle wave functions.  The ions are unobservable, at least until far after the other particles have been recorded, so they do not play a further role in the evolution of the system.  

When the first vector particle decays and is observed, this fixes  the azimuthal direction, and, with it, fix the polarization of the system.  The decay itself does not constitute observation, and so does not collapse the wave function \cite{Klein:2002gc}.  Collapse occurs later, when the decay products are observed.   When the first decay product is observed, it defines
an azimuthal direction, fixing the polarization axis. Essentially, this is like the direction chosen for a polarizing filter  \cite{Baranov:2008zzb}.
The wave function for the remainder of the system is then 
\begin{equation}
\ket{\psi_{n=2}} = \ket{V_2(\theta_1) b(\theta_1)}
\end{equation}
where the  $\theta_1$ is the angle of the first observed vector meson decay. 

When the second particle decays, its polarization is fixed along the axis $\theta_1$, so the decay products will follow the distribution (\ref{eq:DTone}
\begin{equation}
P(\theta_{21}) \propto \cos^2(\theta_{21}) ,
\label{eq:DTQM}
\end{equation}
where $\theta_{21}$ is the angle between the first and second decay.   

\begin{figure}
    \centering
    \includegraphics[width=0.48\textwidth]{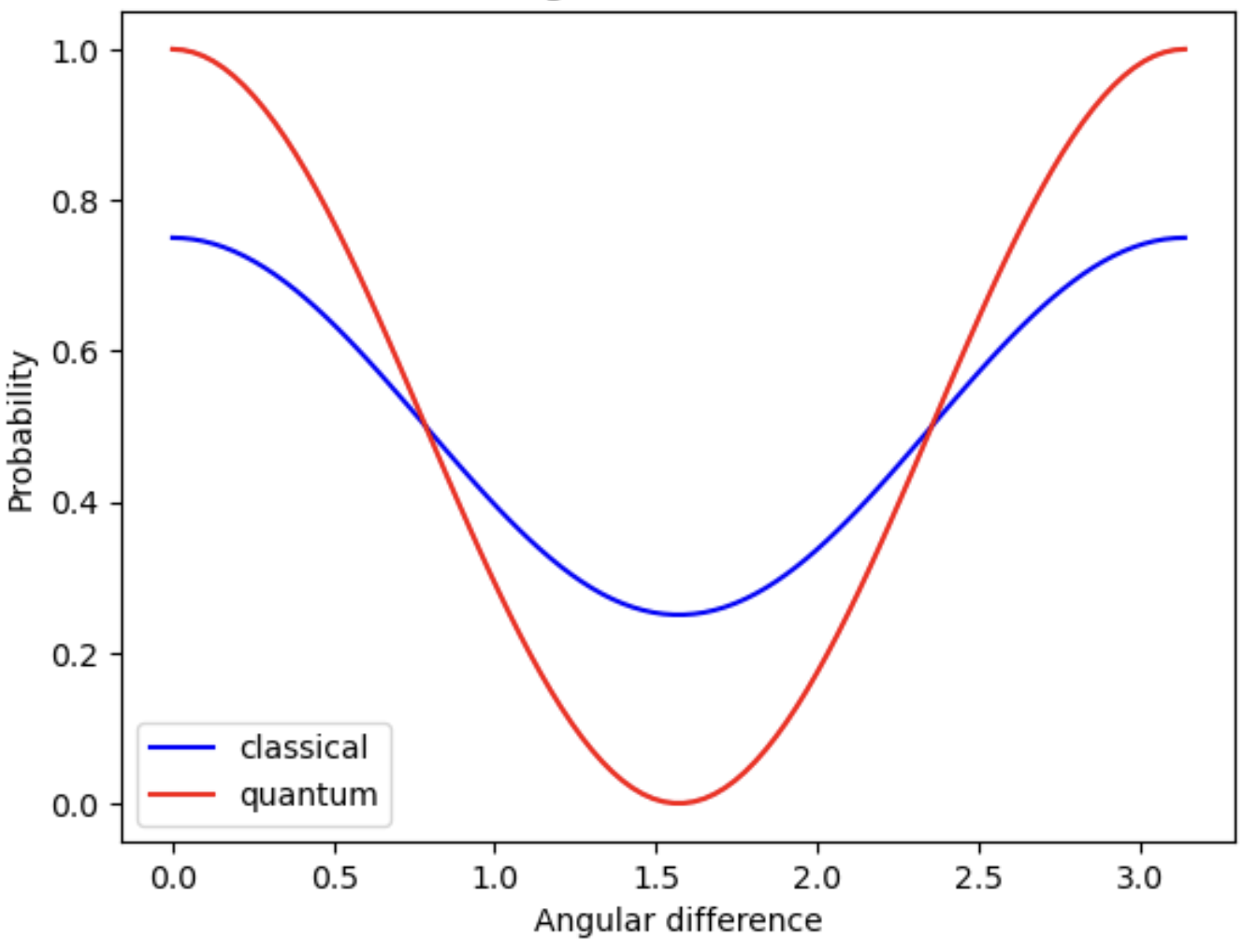}
\caption{Comparison of the classical-calculation and quantum-calculation azimuthal angular distributions for two vector particles produce in a single UPC interaction.}
    \label{fig:angles}
\end{figure}

Figure \ref{fig:angles} compares the classical and quantum predictions for angular correlations, Eqs. 
\ref{eq:DTclass} and Eq. \ref{eq:DTQM} respectively.  The quantum prediction displays a considerably stronger angular dependence, with $P(\pi/2)=0$, in contrast with the classical prediction.  This is similar to the differences between classical and quantum correlations in the Bells inequality, where no light passes through polarizers with an angular difference of $\pi/2$. 

This system has similarities to reactions like $e^+e^-$ production of $\tau^+\tau^-$ \cite{Privitera:1991nz} or $\Lambda\overline\Lambda$ \cite{Tornqvist:1980af}, or production of a pair of vector particles in $\eta_c$ decay \cite{Li:2009rta}.  In these examples, the decay of a first particle selects an axis which sets the polarization of the second.  However, the UPC reactions differ in two ways.  First, the two particles share no common history; the entanglement comes from the symmetry of the system.  And, in UPCs, three or more vector particles can be produced. 

{\it 3 particle correlations.} Final states with three produced vector particles - {\it i. e.} a $\rho$ meson along with mutual nuclear excitation, have already been observed \cite{STAR:2002caw}.  
The $N=3$ initial wave function is a straightforward expansion of the $N=2$ case:
\begin{equation}
\ket{\psi_3} = \ket{V_1(\theta)V_2(\theta)V_3(\theta)b(\theta)} 
\label{eq:three} 
\end{equation}
Each vector particle is entangled with the other two, and the three particles are in a fully inseparable, maximally entangled
Greenberger-Horne-Zeilinger (GHZ) state \cite{Dur:1999re,Braunstein:2005zz,Bouwmeester:1998iz}.  As with the two-vector case, the decay and observation of the first particle sets the initial polarization direction to $\theta_1$: 
\begin{equation}
\ket{\psi_{n=2,3}} = \ket{V_2(\theta_1)V_3(\theta_1)b(\theta_1)}
\end{equation}
As with the two-particle case, when the second particle decay products are observed, they make a $\cos^2(\theta)$ distribution with respect to this direction.  

The third excitation is more interesting, as there are two prior azimuthal directions, $\theta_1$ and $\theta_2$.  Here, one may draw an analogy from photons passing through polarizers, where the photon retains the polarization of the last filter it passed through, irrespective of the settings of the previous filters.  So, the last polarization measurement sets the polarization axis. The third particle to decay will have an azimuthal angular distribution which follows $\cos^2(\theta_{32})$ distribution with respect to the second particle, and a $1+\cos(2\theta_{31})/2$ distribution with respect to the first.  

If there are more than three particles, each successive particle will follow the $\cos^2(\theta)$ distribution with respect to the previous particle.  For large $N$, there is a random walk of the polarization axis as the number of particles increases, similar to the way that one can gradually rotate optical photon polarization with a series of polarizing filters at successively larger angles.

The fact that the average angular separation between two particles depends on the order of observation - essentially the number of intermediate observations between them - highlights the enigmatic role observation plays in quantum mechanics \cite{savitsky2025putting}.  The particle decays themselves cannot constitute an observation, since for $\rho^0$, decay takes place before the amplitudes from the two sources can overlap.  The presence of interference in single vector-meson production shows that the post-decay wave functions must contain amplitudes for all possible decays \cite{Klein:2002gc}.  Instead, observations might occur when the decay products interact with a particle detector - either the first interaction with any material, or an interaction in an active medium that result in a signal being recorded for an analysis.  Neither of these possibilities is fully satisfying. 

In either case, both daughters must be observed.  
The observation of the first daughter should partially collapse the wave function to the subset of phase space that would produce a daughter with those characteristics, while observation of the second will fully determine the polarization direction. For two particles, the time ordering does not matter \cite{Aguilar-Saavedra:2025pnj}, but for three particles, the ordering can depend on the observers frame of reference \cite{castro2012entanglement}.  This is unlikely to be an issue for an experiment where all of the subdetectors are at rest with respect to each other. 

A specific example may be useful:  $\rho$ photoproduction, accompanied by mutual GDR excitation in the ALICE detector at the LHC \cite{ALICE:2012dtf}.  In ALICE, charged particles from $\rho\rightarrow\pi^+\pi^-$ decay are first observed in silicon detectors located about 2.3 cm from the interaction point  \cite{Isakov:2025dqm}.  The neutrons from the GDR decay are detected in Zero Degree Calorimeters (ZDCs) $\pm 116$~m downstream from the interaction point \cite{ALICE:1999edx}.  The ALICE ZDCs are divided into four segments (quadrants), which could provide some azimuthal information.  In ALICE, the vector meson will be observed in the silicon detectors first, followed by the neutrons from GDR decay.  The two neutrons will be observed 2nd and 3rd, so will be more tightly correlated than the average correlation between the vector meson and the two neutrons.  One could also measure multiple vector mesons, at different rapidities (with different path lengths to the detector).   

Although $N>3$ final states are more difficult to observe experimentally,  the large $N$ limit is still of interest.  Due to the gradual random walk, for large $N$, the polarization vectors should approach a random azimuthal angle distribution.

{\it Conclusions.} UPCs are a unique laboratory for studying wave function collapse, since the produced vector particles share a common linear polarization.  UPCs, can also produce systems of three or more entangled particles.   The produced particles are unstable, and the decays are self-analyzing.  For photoproduction of two vector particles, classical and quantum mechanics make very different predictions about the azimuthal angular correlations of the decay products, with the quantum correlations stronger (narrower) than those from classical calculations.  

Systems of three or more vector particles are also possible; they exhibit angular correlations that depend on the order in which the decay products are observed.  For three or more particles, the polarization axis for one particle is set by the decay of the previous particle, leading to a random walk in azimuthal direction for later particles.  Experimental setups with different distances from the production point to the detector elements, might observe different azimuthal angular correlations. 

The cross section to produce two and three vector-particles (GDR excitations or vector mesons)  final states is large. STAR and the LHC experimental collaborations have already collected enough data for some of the two and three particle studies discussed here \cite{Klein:2016dtn,Khatun:2024vgn}. 

The author thanks John Collins for useful discussions, and Sergei Voloshin, Volker Koch and Feng Yuan for looking at the manuscript.   
This work is supported in part by the U.S. Department of Energy, Office of Science, Office of Nuclear Physics, under contract number DE-AC-76SF00098.

\bibliographystyle{apsrev4-1} 
\bibliography{corr}

\end{document}